\documentclass[twocolumn,pre,showpacs,preprintnumbers,amsmath,amssymb,superscriptaddress]{revtex4}

\usepackage{graphicx}% Include figure files
\usepackage{dcolumn}% Align table columns on decimal point
\usepackage{bm}% bold math
\usepackage{longtable}% bold math

\begin{document}

\preprint{APS/123-QED}

\title{L\'evy-type diffusion on one-dimensional directed Cantor Graphs}

\author{Raffaella Burioni}
\affiliation{Dipartimento di Fisica, Universit\`a degli Studi di
Parma, viale G.P.Usberti 7/A, 43100 Parma, Italy}
\affiliation{INFN, Gruppo Collegato di
Parma, viale G.P. Usberti 7/A, 43100 Parma, Italy}
\author{Luca Caniparoli}
\affiliation{Dipartimento di Fisica, Universit\`a degli Studi di
Parma, viale G.P.Usberti 7/A, 43100 Parma, Italy}
\author{Stefano Lepri}
\affiliation{Istituto dei Sistemi Complessi, Consiglio Nazionale delle
Ricerche, via Madonna del Piano 10, I-50019 Sesto Fiorentino, Italy}
\author{Alessandro Vezzani}
\affiliation{CNR-INFM S3, Dipartimento di Fisica, Universit\`a di 
Modena e Reggio Emilia, Via G. Campi 213A, 41000 Modena, Italy}
\affiliation{Dipartimento di Fisica, Universit\`a degli Studi di
Parma, viale G.P.Usberti 7/A, 43100 Parma, Italy}

\date{\today}

\begin{abstract}
L\'evy-type walks with correlated jumps, induced by the topology of the medium, are
studied on a class of one-dimensional deterministic graphs built from
generalized Cantor and Smith-Volterra-Cantor sets. The particle performs
a standard random walk on the sets but is also allowed to move
ballistically throughout the empty regions. Using scaling relations and the 
mapping onto the electric network
problem, we obtain the exact values of the scaling exponents for the asymptotic
return probability, the resistivity and the mean square displacement as a
function of the topological parameters of the sets. Interestingly, the systems
undergoes a transition from superdiffusive to diffusive behavior as a function
of the filling of the fractal. The deterministic topology also allows us to
discuss the importance of the choice of the initial condition. 
In particular, we demonstrate that 
local and average measurements can display different asymptotic behavior.
The analytic results are compared with the numerical solution of 
the master equation of the process.
\end{abstract}

 \maketitle

\section{\label{sec:intro}Introduction}

Stochastic processes characterized by L\'evy walks are relevant in many
physical phenomena, ranging from condensed matter, biological systems and
ecology to transport in turbulent fluids and in porous media. In recent years
there has been a growing interest in the field (see Ref.~\cite{klages} for 
a recent overview). However, 
direct comparison of theoretical predictions with experimental data in 
fully controlled experimental conditions is often unfeasible.
As a matter of fact, evidence of anomalous transport and diffusion
is often indirect and experimental setups allowing a
direct and tunable study of such phenomena are hardly available. A
breakthrough in the field was achieved in a recent experiment, by means of an
engineered material  where light rays were demonstrated to perform a L\'evy
walk-type of diffusion \cite{Levy}. This new class of  materials is built by
packing glass microspheres with diameters following a L\'evy distribution, and
then filling the space between the spheres with strongly scattering
nanoparticles. As light is not scatterered within the microspheres, 
anomalous diffusion arises as a consequence of the power-law
distribution of their diameters. The results of the experiments were indeed
interpreted using a model of
L\'evy walks \cite{Blumen1989,Klafter1990} for photons, and the new material
was named  \emph{L\'evy glass}.  An important feature of the
experimental samples is that  the walk is correlated, and the correlation
is induced by the topology of the quenched medium. Light that has just crossed a
large glass microsphere without being scattered has a high probability of being
backscattered at the subsequent step undergoing a jump
of similar size. Studies on models of L\'evy flights have evidenced that spatial
correlations in jump probabilities have a deep influence on the 
diffusion properties \cite{Fogedby94,Maass,Schulz02}. 
Here, however, we consider L\'evy walk models, where particles either 
move ballistically or are scattered by impurities, and do not perform long range
instantaneous flights. Recently, one-dimensional models for L\'evy walks, where the
correlation is induced by an underlying quenched and correlated random
environment, have been studied. One-dimensional models represent simplified
systems and may not compare quantitatively with real experiments. Nevertheless
they allow  for an exact solution of the dynamics. The recent studies 
focused, respectively, on the mean square displacement
in a L\'evy-Lorentz gas \cite{klafter}, and on the conductivity and 
transmission through a chain of barriers with L\'evy-distributed spacings 
\cite{beenakker}. Interestingly,  in simple random walks  these different aspects are not
independent and they can be connected by assuming standard scaling relations for
the random walk probabilities \cite{cates}, and exploiting the analogy between
the associated master equation and the Kirchhoff equations \cite{doylesnell}. 
When an analogous scaling holds for L\'evy walks, then it can be applied to
relate the scaling exponents of the return probability, the mean square
displacement and the effective resistance of the samples. 
Another subtle point
in random models is represented by the choice of the starting site, as discussed
in Ref.~\cite{klafter}, and it would be interesting to have a direct control on
this problem.

In this paper, we will present a new class of one-dimensional models for
correlated L\'evy walks, that will allow us to investigate in detail the scaling
and the relation between transmission and mean square displacement, as well as 
the dependence on the starting point through average and local diffusion
properties. 
More specifically, we will consider diffusion across deterministic
fractals, namely generalized Cantor and Cantor-Smith-Volterra  sets (the latter
being the simplest examples of fat fractals). The walker dynamics is such that
it performs a standard random walk step on sites belonging to the set, while 
moving ballistically (without changing direction) otherwise.  In the following 
we will  refer to these two types of sites as ``bidirectional'' or
``unidirectional'', respectively. L\'evy-type diffusion may thus arise as the
longer steps will be distributed according to the voids of the Cantor set, much 
in the same spirit of the L\'evy glass experiment \cite{Levy}. 

The model can be mapped onto a random walk on a suitable directed graph  and it
allows to obtain analytical results  by extending the usual scaling arguments. 
In particular,  the scaling of the resistivity with the length of the graph yields the exponents  for the
asymptotics of the return probability and the mean square displacement. 
Interestingly, the system undergoes a transition from superdiffusive to
diffusive behavior as a function of the density of the fractal. A similar
transition has already been evidenced for L\'evy flights in presence of
quenched noise \cite{Maass} and also in the model of Ref.~\cite{beenakker}.

A subtle effect of quenched disorder is the dependence of  the observables on
the choice of the walker starting site \cite{klafter}.  An insight on this issue
is clearly relevant for guiding experimental investigations. Due to the simple
deterministic topology, our model allows for a clear analysis of such issue. In
particular, the choice of a random starting point corresponds to a measure of
quantities averaged over all possible starting sites. On inhomogeneous 
structures, such averages can display a different asymptotic behavior
\cite{burioni} with respect the corresponding local quantities, obtained by 
taking into account a specific starting point. This is indeed the case for
generalized Cantor sets:  scaling relations hold for local quantities but are
violated for average ones. In particular, in the local case, the random walker
shifts  from a superdiffusive to a diffusive motion, as a function of
topological parameters tuning the void density. In the average case the behavior
is even richer:  a change of the void density and size induces a transition
among three different regimes characterized by ballistic, superdiffusive and
standard diffusive motion, respectively.  In this case, which appears to be the
closest to the  experimental settings, adopting a method similar to
Ref.~\cite{klafter}, we give a rigorous bound on the mean square displacement
and discuss the violation of scaling relations. 

The paper is organized as follows: In Sec.~\ref{sec:model} we introduce
the topology and the dynamics of the model. 
In Sec.~\ref{sec:exactloc} we first review the scaling relations connecting
random walk properties and the resistivity and derive the exact results for
the resistivity and for the local asymptotic behavior of the walker itself. We
also check, via the numerical solution of the master equation, the scaling
hypothesis for local quantities. In Sec ~\ref{sec:exactav} we
turn to the average case and provide a rigorous lower bound for the
exponent of the mean square displacement, as a function  of the distribution of
the voids. We show under which condition the bound is fulfilled exactly.
Also in this case  we compare our analytical results with the  numerical
solution of the master equation. Some concluding 
remarks are given in Sec.~\ref{sec:Conclusions}.

\section{\label{sec:model}A random walker on generalized Cantor graphs}

\subsection{Random walk across Cantor Sets}

\begin{figure}
\includegraphics[width=\columnwidth]{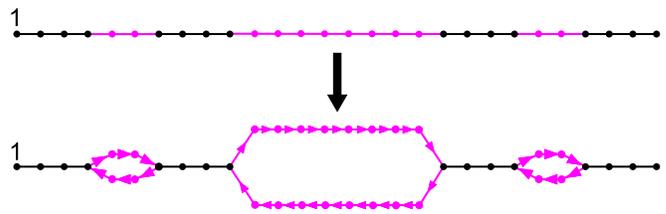}
\caption{(Color online) Upper figure the discrete Cantor set in grey (fuchsia)
the sites which have been removed (voids).
Lower figure the Cantor graph associated to the Cantor 
set at generation $G=3$ where undirectional bubbles have been introduced to describe
ballistic motion in the voids. The site $i=1$ is the origin of the graph.}
\label{cantor_walk_bubble}
\end{figure}

The standard Cantor set is built by removing the middle thirds of a linear chain
and iterating  the rule in every segment. The resulting set has zero measure, 
with fractal (Hausdorff) dimension $d_f^{(set)}=\frac{\log 2}{\log 3}$.
The procedure can be easily generalized:  let
us denote by $n_s$ the number of parts in which we split the segment  ($n_s$
odd). Removing  parts in even positions and iterating the process, the measure
of the remaining set is still zero and its fractal dimension is 
\begin{equation}
d^{(set)}_f \;=\;\frac{\log \frac{n_s+1}{2}}{\log n_s}.
\label{dset}
\end{equation}
If one builds the Cantor set by
growing the fractal, then in this notation $n_r=\frac{n_s+1}{2}$ is the number
of replicas of the generation $G-1$ in the generation $G$. 
The  integer  $n_r$ will be used in the following to
identify a given set. One 
can introduce the discrete version of the above procedure, as in
the upper part of Fig. \ref{cantor_walk_bubble}, where the dark sites represent the Cantor 
set and the grey (fuchsia) sites have been removed and correspond to voids.

Let us now introduce a random walker on such family of Cantor set.
The walker can reside on any site of the lattice.
If it is on the void regions, it moves ballistically
without changing direction: when the random walker enters a void from left
(right), it keeps going left (right) without being scattered until it reaches
the opposite edge of the void itself. If instead the walker is on the sites belonging
to the Cantor set, it performs a standard random walk with $1/2$
probability to move right or left at the next step. 

Denoting with $P_{ij}^\pm(t)$ the probability that a walker
started at site $i$  arrives  at time $t$ at site $j$,
respectively with positive or negative  velocity ($i=1,\ldots,N_{tot}$ can be both a 
site of the Cantor set or a void site, see upper part of Fig.~\ref{cantor_walk_bubble}), 
then the master equation of the corresponding process reads
\begin{eqnarray}
&&P_{ij}^+(t+1) = T_{j-1} P_{ij-1}^+(t) + (1-T_{j-1}) P_{ij-1}^-(t), 
\nonumber \\
&&P_{ij}^-(t+1) = (1-T_{j+1}) P_{ij+1}^+(t) + T_{j+1} P_{ij+1}^-(t).
\label{me}
\end{eqnarray}
The Cantor structure is therefore defined by the transmission coefficients 
\begin{equation}
T_j = 
\begin{cases} 
  1/2,  & \text{if }j\text{ is on the Cantor set} 
  \\
  1, & \text{on the voids}
\end{cases}
\end{equation}
The above choice for $T_j$ yields indeed a ballistic motions on the voids and a random-walk
behavior on the Cantor set. The choice $T_j=T$ is readily recognized to correspond to the standard 
persistent random walk \cite{weiss}.
In this respect, our model can also be regarded as a random walk 
with a site-dependent persistence (see e.g. \cite{Miri2006} and 
references therein). 

\subsection{Cantor Graphs from Cantor Sets}
 
\begin{figure}
\includegraphics[width=\columnwidth]{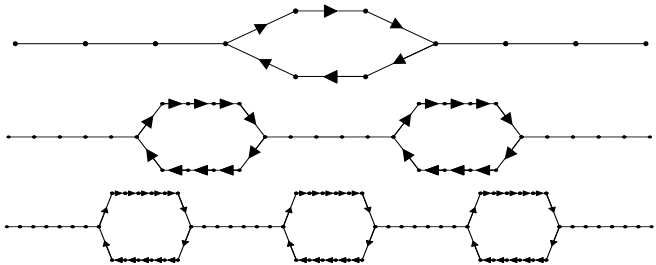}%nat_3-5-7.eps}
\caption{Three examples of Cantor graphs of generation 2 with $n_r=2,3,4$.}
\label{cantor_walk_nat}
\end{figure}

Let us introduce an equivalent approach, based on graph
theory. We associate to the Cantor set a directed \emph{Cantor graph}, obtained
by putting undirected links in every  ``solid'' segment and 
replacing the
voids with bubbles of appropriate length whose sites are connected by
directed links (see Fig.~\ref{cantor_walk_bubble} for an explicit 
construction).
Some examples of
graphs at different $n_r$ are shown in Fig.~\ref{cantor_walk_nat}. In this
framework, the model consists of a random walk on such a directed graph. The
adjacency matrix $A_{ij}$ of the graph is defined as $A_{ij}=A_{ji}=1$ if sites
$i$ and $j$ are connected by an undirected link, $A_{ij}=1$ if the site $i$ is
connected to site $j$ by a link directed from $i$ to $j$ and $0$ otherwise. The
outgoing coordination number $z_i=\sum_j A_{ij}$ is one in the voids and two in
the solid segments. The transition probability for the random walk is therefore:
$p_{ij}=A_{ij}/z_i$.

\begin{figure}
\includegraphics[width=\columnwidth]{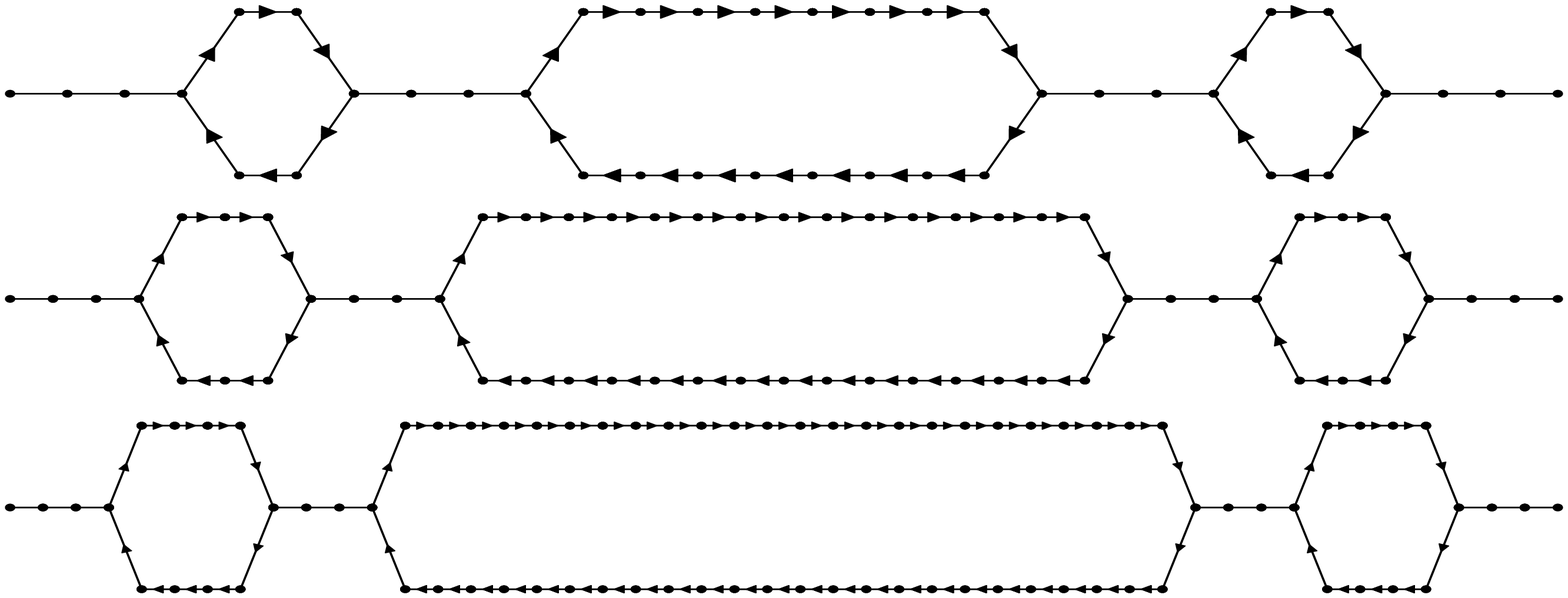}%var_3-4-5.eps}
\caption{The generation $G=3$ of the  graph  with $n_u=3,4,5$, in the 
$n_r=2$ case.}
\label{cantor_walk_var}
\end{figure}

The structure of the voids in the Cantor graph strongly influences the motion of
the random walker, so let us now introduce a further generalization. In the
previous examples, the lengths $L_k$  of the unidirectional segments are 
$L_k=n_s^k$, $k=1,\ldots,G-1$, where $G$ is the generation of the graph.
However, we can choose an arbitrary $n_u$ ($n_u$ integer $\geq 2$), so that  the
length of the unidirectional segments becomes 
\begin{equation}
\label{eq:eq_lungh}
 L_k=n_u^k,\quad k=1,\ldots,G-1,
\end{equation}
as is shown in Fig.~\ref{cantor_walk_var}.

From Eq.~(\ref{eq:eq_lungh}), we can compute $N_b(G,n_r)$ and $N_u(G,n_r,n_u)$, 
the number of bidirectional and unidirectional sites respectively 
(those plotted in black and fuchsia in Fig.~\ref{cantor_walk_bubble}) in a
graph of generation $G$:
\begin{equation}
N_{b}=2n_r^G 
\label{nb}
\end{equation}
\begin{equation}
N_{u} =  
\begin{cases}
\frac{(n_u-1)n_r^G-\ (n_r-1)n_u^G}{n_r-n_u}+1
&\mathrm{if}\ n_u\neq  n_r \\
& \\
- n_r^G+\ (n_r-1)G  n_r^{G-1}+1 & \mathrm{if}\ n_u=  n_r
\end{cases}
\label{nu}
\end{equation}

Accordingly, the total number of sites is $N_{tot}(G,n_r,n_s)$ is
\begin{equation}
\label{ntot}
N_{tot}=
\begin{cases}
\frac{(2n_r-n_u-1)n_r^G-(n_r-1)n_u^G}{n_r-n_u}+1
&\mathrm{if}\ n_u\neq n_r \\
& \\
n_r^G+(n_r-1)Gn_r^{G-1}+1 & \mathrm{if}\ n_u= n_r
\end{cases}
\end{equation}
From Eqs.~(\ref{nb}), (\ref{nu}) and (\ref{ntot}), it follows 
that the measure of both sets $N_b$ and $N_u$ is strictly positive
if  $n_u< n_r$. This choice of the parameters corresponds to the so
called \emph{fat} fractals. An example of this kind of structure is the
Smith-Volterra-Cantor set. Conversely, 
we will name \emph{slim} fractals the graphs with 
$n_u > n_r$, where the measure of the solid segments is vanishing.

It is important to notice that, even if the Cantor set has a fractal dimension
$d_f^{(set)}<1$, Eq.~(\ref{dset}), 
the directed Cantor graph has always a fractal dimension $d_f=1$.
Indeed, since the voids have been filled with one-dimensional bubbles, 
the number of sites within a distance $r$ from a given site  grows 
as $\sim r^{d_f}$, with $d_f=1$.

\section{\label{sec:exactloc}Diffusion on Cantor Graphs: Local Behavior}

\subsection{Asymptotic properties of random-walks, spectral dimensions and Einstein relations}

On the directed Cantor graph, we can now define the standard random walks
quantities. Their asymptotic behavior are expected to be described by  power
laws, with characteristic exponents, usually related only to the large-scale
topology of the structure \cite{burioni}. 

Let $P_{ij}(t)$ be the probability that a walker started at 
site $i$ arrives at $j$ in $t$ steps, $P_{ij}(t)=P_{ij}^+(t)+P_{ij}^-(t)$.
In particular, $P_{ii}(t)$ is the random walk
autocorrelation function i.e. the return probability to the starting
point after $t$ steps. At large times $t$ it is expected that \cite{Orbach}
\begin{equation}
P_{ii}(t) \sim t^{- {d_s\over 2}}
\label{ds}
\end{equation}
where $d_s$ often coincides with the spectral dimension, which also rules the
low eigenvalues region of the spectrum for the discrete Laplacian on the graph
\cite{debole}. In the case at hand, being the graph directed, $d_s$ simply describes the
asymptotic behavior of the return probability, but we will refer to it as a
"generalized" spectral dimension as well. 

The mean square displacement from the starting site $i$ after $t$ steps is:
\begin{equation}
\langle x^2_i \rangle - \langle x_i \rangle^2 \equiv \left( \sum_j x^2_{ij} P_{ij}(t)\right)-
\left( \sum_j x_{ij} P_{ij}(t)\right)^2.
\label{spostamentoq}
\end{equation}
where $x_{ij}$ is the distance between the sites $i$ and $j$.  If the starting site corresponds
to the origin of the Cantor graph ($i=1$ in Fig.  \ref{cantor_walk_bubble}) and periodic boundary 
condition are chosen, as in all our numerical solutions, then $\langle x_i \rangle=0$ 
for simmetry reasons. The exponent $\gamma$ defined by the asymptotic behavior
\begin{equation}
\langle x^2_i \rangle \sim t^{\gamma}
\label{spostamentoqasint}
\end{equation}
classifies the diffusive properties of the random walker: $\gamma=1$ 
corresponds to usual diffusion, $\gamma > 1$ to superdiffusion, and $\gamma<1$ is
typical of a subdiffusion. The value $\gamma=2$ characterizes a ballistic motion
where the random walker displacement grows linearly with time. 

It is well known that on regular structures, the dynamic exponents ruling the
return probability and the mean square displacement are not independent, i.e.
for systems where scaling holds \cite{cates}, 
\begin{equation}
\gamma=d_s/d_f
\label{gamma}
\end{equation}
In the Appendix we will show that such a relation still holds under a 
generalized scaling hypothesis that takes into account the underlying 
fractal structure of the Cantor graph. More precisely, we assume 
that $P_{ij}(t)$ depends on $j$ only through the distance $x_{ij}\equiv r$,
between $i$ and $j$, i.e. $P_{ij}(t)\equiv P_i(r,t)$ and that
\begin{equation}
P_{i}(r,t)\;=\;t^{-d_s/2} f_i\left(\frac{r}{\ell(t)},g(\log_x \ell(t))\right)
\label{scalhyp}
\end{equation}
where $\ell(t)$ is the correlation length of the system and 
$g(\cdot)$ is a periodic function of a logarithm in arbitrary base $x$.
The master function $f_i$ can depend on the initial site. 
The scaling factor $t^{-d_s/2}$
ensures consistency with the definition of the return probability, 
Eq.~(\ref{ds}), as seen by setting $j=i$ ($r=0$) in Eq.~(\ref{scalhyp}).
Moreover, log-periodic oscillations 
are superimposed to the leading exponential behavior 
as it is expected on fractals \cite{woess}.
In the Appendix we also show that, up to log-periodic corrections,
$\langle x_i^2\rangle \sim \ell^2(t)$ and that
\begin{equation}
\ell(t) = t^{\frac{d_s}{2 d_f}}.
\label{lt} 
\end{equation}
In general $P_{i}(r,t)$ presents the above scaling form for  large enough times.
In our calculations we verified that if the walker starts from the origin of the
structure, scaling is realized in a few steps. A different choice can
give rise to very long transients where $P_{ij}(t)$ may also depend on
the direction and not only on the distance $r$ alone. The net 
effect is a nonvanishing drift $\langle x_i (t)\rangle\not = 0$ for 
short times.

A simple analogy between the master equation of the random walk and the
Kirchhoff equations \cite{doylesnell} allows to associate to each directed
graph a networks of resistors (see again the Appendix). In this framework, 
it is possible to compute analytically the exponent $\alpha$ 
describing the growth of the resistance $\Omega$
as a function of the distance $r$ between contacts, i.e. $\Omega \sim r^{\alpha}$.
Exploiting
Eq.~(\ref{scalhyp}), one can prove also the following Einstein
relation 
\begin{equation}
\alpha = \frac{2d_f}{d_s}-d_f
\label{einst}
\end{equation}
For local quantities and exponents, we can thus use Eq.~(\ref{einst}) to
calculate $d_s$. We will verify it numerically, also testing the validity of
the scaling  hypothesis. 

\subsection{Local spectral dimension}

When scaling holds, the local spectral dimension can be obtained using the
scaling law of the resistance $\Omega$ and the relation (\ref{einst}). In the
network of resistors associated to the Cantor graph  every bidirectional link
has resistance 1, while for unidirectional  links, a \emph{whole} bubble has
unit resistance (see Appendix).  The resistance between the extremes of the
structure of generation $G$ can be evaluated using the recurrence relation:
\begin{displaymath}
  \Omega (G,n_r)=n_r\ \Omega(G-1,n_r)+(n_r-1).
\end{displaymath}
Using the initial condition $\Omega (1,n_r)=2n_r-1$, one has
\begin{equation}
\label{uwopoccc}
  \Omega (G,n_r) \;=\;2 n_r^{G}-1.
\end{equation}
Expressing $\Omega$ as a function of the distance $r$ between contacts   
at generation $G$, in the $G\to\infty$ asymptotic limit, we obtain the exponent $\alpha$
and, via the scaling relation (\ref{einst}), the value of $d_s$, ruling the return probability and the mean 
square displacement:
\begin{equation}
\label{eq:d_s}
  d_s=\left\{\begin{array}{lc}
\frac{2}{1+\frac{\log n_r}{\log n_u}}, & \mathrm{if}\ n_u> n_r\\
\\
\frac{2}{2-\frac{\log \log r}{\log r}}, & \mathrm{if}\ n_u= n_r
\\
\\
1 & \mathrm{if}\ 
n_u< n_r
\end{array}\right.,
\end{equation}
and $\gamma=d_s$, where we have used that $d_f=1$.
Interestingly, the system exhibits a transition from a superdiffusive to a normally diffusive 
regime at $n_u=n_r$. The random walker on the slim Cantor graphs experiences
a superdiffusive regime while it presents a normal diffusion on fat structures, 
with logarithmic corrections in the critical case $n_u=n_r$. Intuitively, slim Cantor 
graphs mainly consist of long unidirectional bubbles, 
which lead to a superdiffusive behavior, while in the 
opposite case the bubbles practically disappear and hence 
one could expect a normal diffusion. 

\subsection{Numerical results: The master equation and generalized scaling}

For a simple and effective numerical study of the system 
we solved iteratively the master equation Eq.~(\ref{me}) with initial 
condition $P^{\pm}_{ij}(0)=\delta_{ij}/2$ and periodic boundary conditions.

\begin{figure}
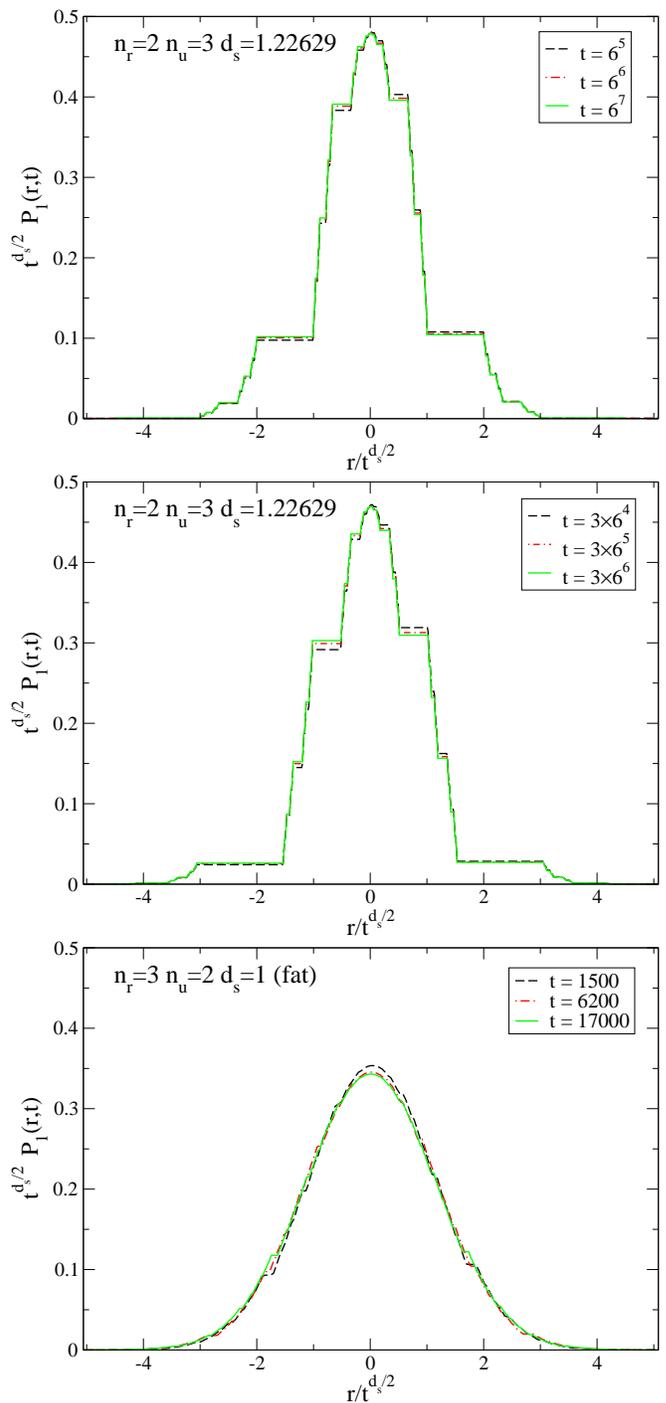

\includegraphics[width=\columnwidth,clip]{pscaling_33.eps}
\includegraphics[width=\columnwidth,clip]{pscaling_33_tdiversi.eps}
\includegraphics[width=\columnwidth,clip]{pscaling_52_tdiversi.eps}
\caption{Dynamical scaling of the probability for initial 
site $i=1$. The size of the lattices correspond to generations $G=9,8$ 
respectively.} 
\label{dynscal}
\end{figure}

We first tested the dynamical scaling of the probability density (\ref{scalhyp}), considering for $g(\cdot)$ a periodic function with unit period in $\log_{n_u}(\ell(t))$, so that
the log-periodicity takes into account the self similarity of the underlying
structure. It is thus convenient to look at the 
data at constant $g(\log_{n_u}\ell(t))$, namely, as prescribed by 
Eqs.~(\ref{lt}) and (\ref{eq:d_s}), at successive times 
\begin{equation}
\label{reltime}
t_k = t_0 \left[n_u n_r\right]^k .
\end{equation}
Figs. \ref{dynscal} refers to a slim and a fat structure.
For the former case the scaling is reasonably accurate,
although the convergence to the asymptotic shape is not 
complete. Moreover the two panels show that for different sequences of times 
satisfying relation (\ref{reltime}) (i.e. different $t_0$) the scaling function
changes, evidencing that the introduction of a generalized scaling 
with a log-periodic term is necessary in this situation. 
The scaling function also depends on the starting site $i$, while 
the exponent are site independent.
For the fat case, the curves tend to be closer and 
closer to a Gaussian, confirming that the diffusion 
is normal. Times have not to be chosen according 
to Eq.~(\ref{reltime})
since in this case log-periodic oscillations are not present.

\begin{figure}
\includegraphics[width=\columnwidth,clip]{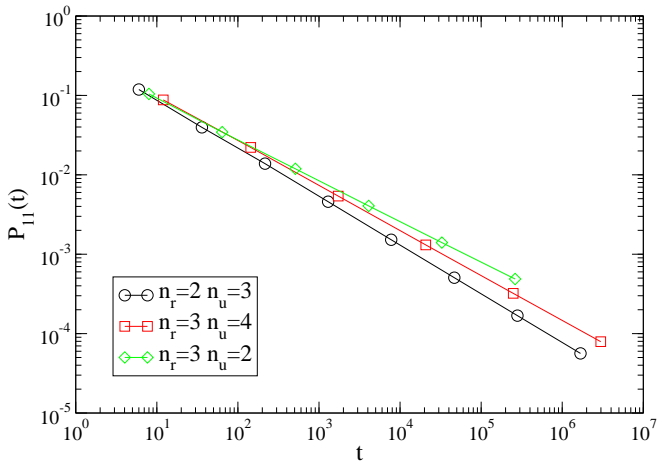}
\caption{Dynamical scaling of the probability for initial 
site $i=1$. Lattices of size $N_{tot}$ correspond to $G=9,7,8$ 
respectively.
To avoid the big jumps and zeros  
what is actually plotted is a coarse grained probability
$\sum_{t=t_{k-1}}^{t_k} \, P_{ii}(t)/[t_k-t_{k-1}]$.
} 
\label{p11}
\end{figure}

\begin{figure}
\includegraphics[width=\columnwidth,clip]{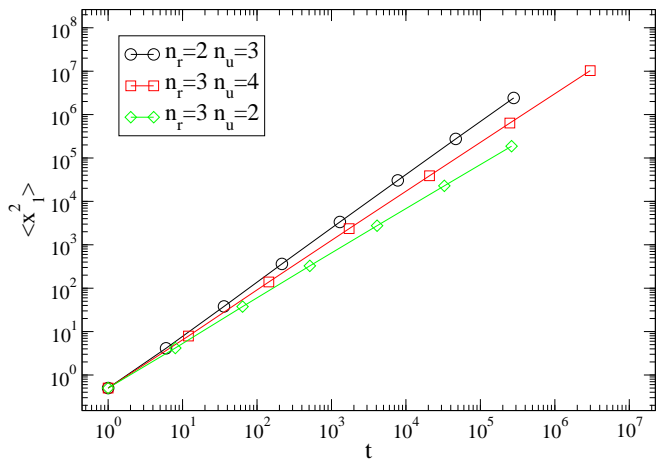}
\caption{Growth of the mean square displacements for initial 
site $i=1$. Lattices of size $N$ corresponding to $G=9,8,7$ 
respectively.} 
\label{x2}
\end{figure}

Figs.~\ref{p11} and \ref{x2} demonstrate that the probability of return 
$P_{ii}(t)$ and the mean square displacement 
behave asymptotically (up to log-periodic corrections) as
prescrived by Eqs.~(\ref{ds}), (\ref{spostamentoqasint}) 
and (\ref{eq:d_s}) ($d_f$=1 for all Cantor graphs). Indeed, 
power-law fits of the curves give exponents $1.21, 1.13, 1.03$
which are in excellent agreement with the theoretical values 
of $d_s$, Eq.~(\ref{eq:d_s})($d_s=1.2262 \ldots, 1.11577 \ldots, 1$ 
respectively).
 
\section{\label{sec:exactav}Diffusion on Cantor graphs: average behavior}

The exponents characterizing the asymptotic behavior of random walks are 
in general independent of the starting site $i$, even on inhomogeneous structures . However, due
to inhomogeneity,  averages over the starting points can result in radically
different asymptotic behaviors with respect to local case \cite{burioni}. 
It is thus necessary to introduce some further averaging on the 
initial conditions.
Let $S_{k}$ ($k$
integer) be a sequence of subgraphs ($S_{k}\subset S_{k+1}$) covering the
infinite graph. 
We define the average mean square displacement:
\begin{equation}
\overline{\langle x^2 \rangle} = 
\lim_{k \to \infty} {1\over N_k} \sum_{i \in S_k} \langle x^2_i \rangle
\label{spostamentomedio}
\end{equation}
($N_k$ is the number of sites in $S_k$) and the average return probability
\begin{equation}
\overline{P(t)} =\lim_{k \to \infty} {1\over N_k} \sum_{i \in S_k} P_{ii}(t)
\label{Pmedio}
\end{equation}
Interestingly, due to the fact that the thermodynamic limit does not 
commute with the large time limit, 
the average quantities (\ref{spostamentomedio},\ref{Pmedio}) 
can feature a different asymptotic behavior with respect to their 
local counterparts. Therefore, as done for the local case, Eqs.~(\ref{ds}) and 
(\ref{spostamentoqasint}), we introduce 
the exponents $\bar{d}_s$ and $\bar \gamma$ for the average quantities:
\begin{equation}
\overline{P(t)} \sim t^{- {\bar d_s\over 2}}, \quad
\overline{\langle x^2 \rangle}  \sim t^{\bar \gamma}.
\label{dsm}
\end{equation}

On the Cantor graph, average exponents not only differ from the local ones,
but also they do not satisfy relations (\ref{gamma}) and (\ref{einst}). 
Indeed, for average quantities scaling is violated and 
the topology of the graph influences diffusion in a
highly non trivial way. Diffusion is ballistic on slim ($n_u>n_r$) Cantor
graphs, while on fat graphs, even if $d_s$ always equals one, there are two different 
scenarios:
\begin{equation}
\label{eq:bargamma}
  \bar \gamma=\left\{\begin{array}{lc}
3-\frac{\log n_r}{\log n_u}, & \mathrm{if}\ n_u< n_r<n_u^2
\\
\\
1 & \mathrm{if}\ 
n_u^2< n_r
\end{array}\right.,
\end{equation}
In both cases, we obtain a lower bound for the
average mean square  displacement  and then we check numerically 
that this lower bound is  satisfied when  $n_r<n_u^2$, obtaining
the exponent given in Eq.~(\ref{eq:bargamma}).

In analogy with \cite{klafter},
let us study  the probability $P_b(l)$ to take a first ballistic step of length $> l$. 
We will show that, on slim Cantor graphs, this
probability approaches $1$ when $G\to\infty$ for every $l$, 
and the motion is therefore
ballistic on average. On the other hand, on fat Cantor graphs,
the limit is a finite number going to $0$ for $l \to \infty$ with a
characteristic power law.
Let us consider the unidirectional segments of length greater than $l$: the 
``favorable'' sites in a unidirectional segment of length $L_k>l$ are $L_k-l$, 
and we have to sum over $k$.
Recalling that the number of segments of length $L_k=n_u^k$ is $(n_r-1) n_r^{G-1-k}$,
we can write $P_b(l)$ as
\begin{equation}
\label{eq:sdasa}
  P_b(l)=\lim_{G\to\infty}
\frac{\sum_{k=k_0}^{G-1}
\left(n_u^k-1 -l\right)\ (n_r-1)\left(n_r\right)^{G-1-k}}
{N_{tot}(G,n_s,n_u)},
\end{equation}
where $k_0$ is an integer chosen such that $n_u^{k_0} > l$, i.e. $k_0:=\left\lceil \frac{\log (l)}{\log n_u}\right\rceil$
(reminding that $\lceil x \rceil:=\min\{n\in \mathbb{Z}\ |\ n\geq x \}$ is the
ceiling function), and $N_{tot}(G,n_r,n_u)$ is the total number of sites in 
the graph of generation $G$, given in (\ref{ntot}). 
The series in the numerator of (\ref{eq:sdasa}) can be summed and we obtain:
\begin{multline*}
  P_b(l)=\lim_{G\to\infty}\frac{1}{N_{tot}}\Bigg[\frac{n_r-1}{n_u-n_r}n_u^G\\
+\bigg[\frac{n_r-1}{n_r-n_u}\left(\frac{n_u}{n_r}\right)^{k_0}-(1+l)n_r^{-k_0}
\bigg]n_r^G+(1+l)\Bigg].
\end{multline*}
Therefore, we have two exponential terms with different bases:
the leading term for $G\to \infty$ will then be different depending on $n_r$ 
and $n_u$.

If $n_u>n_r$, ( i.e. on a slim Cantor graph), considering the asymptotic behavior of 
$N_{tot}(G,n_r,n_u)$ in Eq.~(\ref{ntot}), one obtains
\begin{displaymath}
  P_b(l)=\lim _{G\to \infty}
\frac{-n_u^G(n_r-1)+o(n_u^G)}{-n_u^G(n_r-1)+o(n_u^G)}=1.
\end{displaymath}
Thus, the probability of going through a ballistic step of length 
$>l$ is $1$ for every $l$ and the walker behaves ballistically on average.

Let us now consider the same limit in the case $n_u < n_r$. 
Omitting the ceiling function in $k_0$ (it can be done with a suitably chosen $l$), we 
have
\begin{equation}
 P_b(l) \sim \left(\frac{n_r}{n_u}\right)^{-k_0} =l^{1-\frac{\log n_r}{\log n_u}}
 \label{pbal}
\end{equation}

\begin{figure}
\includegraphics[width=\columnwidth,clip]{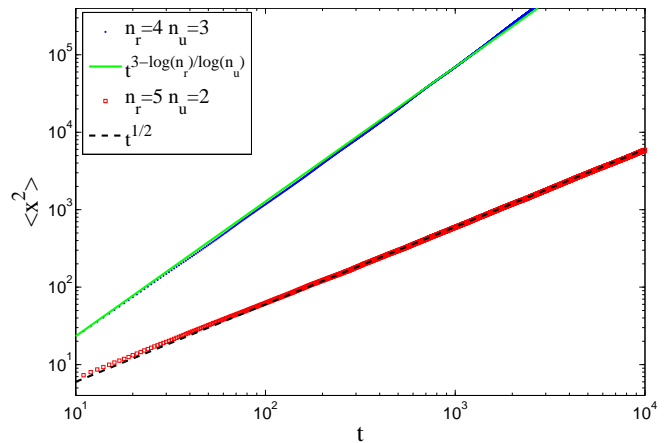}
\caption{Average mean square displacement, as defined in
Eq.~(\ref{spostamentomedio}) for the fat fractals with $n_r=4$, $n_u=3$
and  $n_r=5$, $n_u=2$. Solid and dashed lines are the superdiffusive,  Eq.~(\ref{bound}),
and diffusive behaviour expected in the two structures, respectively.} 
\label{x2fat}
\end{figure}

In order to estimate the average mean square displacement 
in the fat case, let us introduce the average probability of being 
at distance $r$ after $t$ steps as $\bar P(r,t)$. It is convenient 
to split it in two parts writing
\begin{equation}
\bar P(r,t)=\bar P^*(r,t; r<t)+ \frac{P_b (t)}{2} \left(\delta(r-t)+\delta(r+t)\right),
\label{split}
\end{equation}
where $P_b$ is the average probability of performing $t$ consecutive 
ballistic steps and $\bar P^*(r,t; r<t)$ is the average probability of 
arriving in $t$ step at $r$  after some scattering. 
Eq.~(\ref{split}) provides a lower-bound for the mean square 
displacement 
\begin{equation}
\begin{split}
\overline{\langle x^2 \rangle} & = \int_0^\infty dr \ r^2 \bar P(r,t)\\
& > \int_0^\infty dr \ r^2 \frac{P_b (t)}{2} \left(\delta(r-t)+\delta(r+t)\right) \sim t^{3-\frac{\log n_r}{\log n_u}}.
\label{bound}
\end{split}
\end{equation}
where we used expressions (\ref{pbal}) for $P_b(t)$.
For $1<\frac{\log n_r}{\log n_u}<2$ the inequality (\ref{bound}) proves that 
the system is superdiffusive and 
$\overline{\langle x^2 \rangle} \sim t^{3-\frac{\log n_r}{\log n_u}}$
is expected to be the correct asymptotic behavior of the average mean square displacement, 
In Fig.~\ref{x2fat} we compare our analytical prediction with numerical data, 
showing an excellent agreement \cite{nota}. For 
$\frac{\log n_r}{\log n_u}>2$ inequality (\ref{bound}) is trivial 
and a normal diffusion is expected, since the ballistic stretch 
does not provide a significant contribution to 
$\overline{\langle x^2 \rangle}$. The presence of normal diffusion
in this case is also evidenced in Fig.~\ref{x2fat}.

\begin{figure}
\includegraphics[width=\columnwidth,clip]{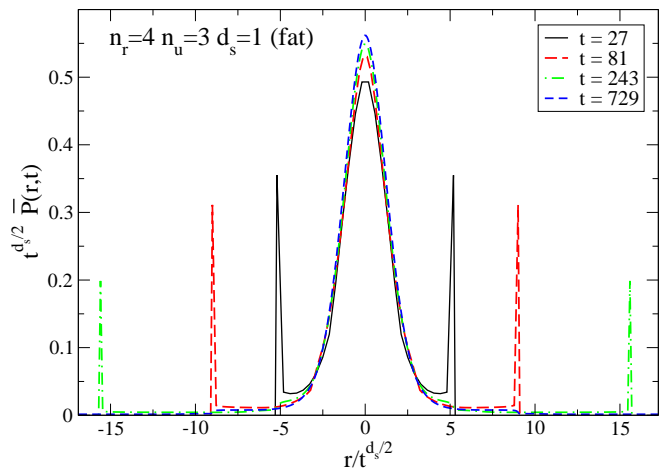}
\caption{Dynamic scaling of the averaged probability $\bar P$
with $n_r=4$ and $n_u=3$ lattices of size $N_{tot}=58976$ ($G=7$).
The average is on all initial sites.} 
\label{PrtFAT}
\end{figure}

Fig.~\ref{PrtFAT} shows that, in the fat case, 
there are actually two different 
contributions to $\bar P(r,t)$:  a ballistic peak lowering with $t$ and a 
central peak which scales as in normal diffusion. This behavior clearly 
breaks the scaling hypothesis (\ref{scalhyp}) and relations (\ref{gamma}) and 
(\ref{einst}) 
between exponents do not hold. Indeed, for $\log(n_r)/\log(n_u)<2$, 
$\bar{\gamma}={3-\frac{\log n_r}{\log n_u}}$, while $\bar d_s=1$ and $\bar \alpha=1$.

\section{\label{sec:Conclusions} Conclusions}

The recent experiments on light scattering in disordered materials open the way
to direct and tunable measurements on systems performing L\'evy walks. The
quenched disorder in the experimental samples has been engineered in order to
obtain a specific effects, and the geometry induces a correlation in L\'evy
walks. In this paper, we take a first step in understanding the influence of
geometry in these disordered samples by studying a one dimensional structure
obtained from the generalization of Cantor and Cantor-Smith-Volterra sets, where
a random walker performs steps that are both L\'evy distributed and correlated. 
Although the applicability of such idealized one-dimensional models to real
experiments is not granted, much theoretical insight can be obtained from their
solution. Using an exact mapping to an equivalent directed graph for the
process, we determine the asymptotic behavior for the return probability and the
mean square displacement and study the validity of the Einstein relations in the
local and average case. 

From our results, it appears that the average case is the closest to the real
experimental setting  \cite{Levy}, and further analysis in higher dimensions are
currently under consideration. An important point is that experimental
samples do contain a certain degree of disorder, which must be taken into
account in order to understand to what extent it can affect diffusion
properties. 

\appendix*
\section{} 

In this Appendix we show that the Einstein relation Eq.~(\ref{einst}) 
between different exponents is a 
direct consequence of the generalized scaling hypothesis, 
Eq.~(\ref{scalhyp}).

Since $P_{ij}(t)$ is a probability, its sum over all sites $n$ is normalized to
one at every time $t$. Moreover, in the scaling hypothesis, all the 
spatial dependence is encoded by the distance $x_{ij}\equiv r$, therefore 
such a sum can be evaluated by introducing an integral in $Kr^{d_f-1}~dr$
where $d_f$ is the fractal dimension and $K$ is a suitable constant, obtaining
\begin{equation}
t^{-d_s/2}\,
\int f_i\left(\frac{r}{\ell(t)},g( \log_x \ell(t))\right) K r^{d_f-1} dr
\;=\; 1.
\label{E2}
\end{equation}  
Changing the integration variable to $r/\ell(t)$ we have
\begin{equation}
t^{-d_s/2}\;\ell^{d_f}(t)\;G(t)  \;=\; 1
\label{E3}
\end{equation}  
where $G(t)$ is a log-periodic function. Eq.~(\ref{E3}) is satisfied 
if  $G(t)=1$ and if Eq.~(\ref{lt}) holds. 

To derive Eq.~(\ref{gamma}), let us consider the mean square displacement:
\begin{equation}
\langle x^2_i(t)\rangle = t^{-d_s/2}\,
 \int r^2 f_i \left(\frac{r}{\ell(t)},g( \log_x \ell(t))\right) K r^{d_f-1} dr
%\label{E4}
\end{equation} 
Upon choosing again the new integration variable $r/\ell(t)$, we have 
\begin{equation}
\langle x^2_i \rangle = \ell(t)^2\;G_1(t)=\; t^{d_s/d_f}\;G_1(t)
%\label{E4}
\end{equation} 
where $G_1(t)$ is a log-periodic function.

Let us finally consider the equation for the electric potential $V_i$ on a 
network of unitary resistors where a unitary current flows from site $0$ to
site $n$, we have:
\begin{equation}
-\sum_j L_{ji} V_j = \delta_{i0}-\delta_{in}
\label{elect}
\end{equation} 
where $L_{ij}=z_i \delta_{ij}-A_{ij}$ is the Laplacian matrix.
Notice that according to (\ref{elect}), in  a directed Cantor graph the inner 
links of a bubble are short-circuit, and a whole bubble has resistance one.
In the framework of random walks on directed Cantor Graph
the master equation (\ref{me}) can be recasted as:
\begin{equation}
P_{0i}(t+1) -P_{0i}(t) = -\sum_j L_{j,i} P_{0j}(t)/z_j + \delta_{i0}\delta_{t0}.
\label{rweq}
\end{equation} 
Denoting with $\tilde{P}_{0i}(\omega)$ the Fourier transform of $P_{0i}(t)$  
we get
\begin{equation}
\tilde{P}_{0i}(\omega)(e^{i\omega}-1) = -\sum_j L_{j,i} \tilde{P}_{0j}(\omega)/z_j + \delta_{i0}
\label{rweq2}
\end{equation} 
comparing Eqs.~(\ref{elect}) and (\ref{rweq2}) 
\begin{equation}
V_i = \frac{1}{z_i}\lim_{\omega \to 0} (\tilde{P}_{0i}(\omega) - \tilde{P}_{ni}(\omega))
%\label{E4}
\end{equation}
The potential difference between 
sites $0$ and $n$ as a function of their distance $r$, can be, hence, 
obtained  introducing in $\lim_{\omega \to 0}\tilde P_{0n}(\omega)$
the scaling relation Eq.~(\ref{scalhyp})
\begin{equation}
V(r) \sim \lim_{\omega \to 0}\int\; e^{i \omega t} t^{-d_s/2}
f_i \left(\frac{r}{\ell(t)},g( \log_{x}\ell(t))\right)
\label{E5}
\end{equation}
Changing the variable of integration into $t'=\omega t$ we get
\begin{equation}
V(r) \sim r^{2d_f/d_s-d_f} 
\lim_{\omega \to 0} G_3\left (\frac{r}{\tilde \ell(\omega)},
g(\log_{x}\tilde\ell(\omega))\right)
\label{E6}
\end{equation}
where $G_3$ is a suitable function and $\tilde{\ell}(\omega)=
\omega^{-d_s/(2d_f)}$ is the correlation length in terms of the frequency 
$\omega$. Therefore, Eq.~(\ref{einst}) holds.

\acknowledgments

We acknowledge useful discussion with D. ben-Avraham,
P. Barthelemy, J. Bertolotti, R. Livi, D.S. Wiersma, K. Vynck.
This work is partially supported by the CNR RSTL project 
N. 827 \textit{Dinamiche cooperative in strutture 
quasi uni-dimensionali}.


\begin{thebibliography}{}
\bibitem{nota}
In Fig.~\ref{x2fat} we introduce a correction in order to evaluate averages on
infinite graphs. Let $\overline{\langle x^2 \rangle}_{G}$ be the mean square
displacement evaluated at the generation $G$. Then we have $\overline{\langle x^2
\rangle}=\overline{\langle x^2 \rangle}_{G}(1-P_{G})+t^2 P_{G}$, where $P_{G}$
is the probability of being at time $t=0$  in a bubble
larger than the largest bubble of $G$ on the infinite graph . Clearly we expect that such a correction
is valid only for time much smaller than the size of this largest bubble. Indeed
only in this case the average motion for walkers starting on such bubbles can be
considered purely ballistic.
\end{thebibliography}
\end{document}